\documentclass[letters,fleqn,usenatbib]{mnras}

\usepackage{newtxtext,newtxmath}
\usepackage[T1]{fontenc}
\usepackage{ae,aecompl}

\usepackage{graphicx}	% Including figure files
\usepackage{amsmath}	% Advanced maths commands
\usepackage{amssymb}	% Extra maths symbols
\usepackage{booktabs}   % Creates booktab tables
\usepackage{multirow}   % Enables vertically merged cells in tables 
\usepackage[normalem]{ulem} % Only for strikethrough text

\title[Compact groups in EAGLE]{Abundance and group coalescence timescales of compact groups of galaxies in the EAGLE simulation}

\author[L. Hartsuiker et al.]{
Len Hartsuiker,$^{1}$\thanks{E-mail: lenhartsuiker@gmail.com}
Sylvia Ploeckinger$^{1,2}$\thanks{E-mail: ploeckinger@strw.leidenuniv.nl}
\\
$^{1}$Leiden Observatory, Leiden University, P.O. Box 9513, 2300 RA Leiden, the Netherlands\\
$^{2}$Institute for Computational Cosmology, Department of Physics, University of Durham, South Road, Durham, DH1 3LE, UK
}

\date{Accepted XXX. Received YYY; in original form ZZZ}

\pubyear{\the\year}

\DeclareRobustCommand{\DE}[3]{#2} % set up for citation

\begin{document}
\label{firstpage}
\pagerange{\pageref{firstpage}--\pageref{lastpage}}
\maketitle

\begin{abstract}
Observations of compact groups of galaxies (CGs) indicate that their abundance has not significantly changed since $z=0.2$. This balance between the timescales for formation and destruction of CGs is challenging if the typical timescale for CG members to merge into one massive galaxy is as short as historically assumed ($<$ 0.1~Hubble times). Following the evolution of CGs over time in a cosmological simulation (EAGLE), we quantify the contributions of individual processes that in the end explain the observed abundance of CGs.
We find that despite the usually applied maximum line-of-sight velocity difference of $1000\,\mathrm{km\,s}^{-1}$ within the group members, the majority of CGs ($\approx 60$ per cent) are elongated along the line-of-sight by at least a factor of two. These CGs are mostly transient as they are only compact in projection. In more spherical systems $\approx 80$ per cent of galaxies at $z>0.4$  merge into one massive galaxy before the simulation end ($z=0$) and we find that the typical timescale for this process is 2-3~Gyr.
We conclude that the combination of large fractions of interlopers and the longer median group coalescence timescale of CGs alleviates the need for a fast formation process to explain the observed abundance of CGs for $z<0.2$.
\end{abstract}


%%%%%%%%%%%%%%%%%%%%%%%%%%%%%%%%%%%%%%%%%%%%%%%%%%

%%%%%%%%%%%%%%%%% BODY OF PAPER %%%%%%%%%%%%%%%%%%

\section{Introduction}
Compact groups of galaxies (CGs) are the densest galaxy systems known. Galaxies that are part of a CG are only separated by projected distances in the sky of a few tens of kiloparsecs.
As CGs are so dense, they provide a very interesting environment for the study of galaxy interactions and evolution. This has frequently lead to the creation of CG samples to study their properties (starting with e.g. \citealp{turner}; \citealp{hickson1977}; \citealp{heiligman}). 
Arguably the most famous sample was created by \citet{Hickson}, who systematically searched for CGs using criteria on the isolation and the compactness of galaxy groups on the sky.
These criteria or slightly modified ones are still regularly used to create CG catalogs (e.g. \citealp{prandoni1994}; \citealp{lee2004}; \citealp{decarvalho2005}; \citealp{mcconnachie}; \citealp{diaz2012}; \citealp{diaz2018}).

A recent catalogue of \citet{sohn2015} revealed that the completeness corrected abundance of CGs is constant within an order of magnitude for $z<0.2$. This is unexpected if the estimated crossing times of CGs are very short ($\approx$0.02 Hubble times, \citealp{barnes1989}) and the theoretical timescales for all CG members to spiral into the group center by dynamical friction are of order a few crossing times, as historically assumed (\citealp{whiterees1978}).
If all galaxies in a CG merge into a single galaxy in this short time span, new CGs would have to form within a similarly short time (e.g. within rich groups, \citealp{diaferio1994}) to explain their abundance. 

Alternative explanations include projection effects, long coalescence timescales, or a combination of both. 
If a large fraction of CGs only appear to be compact in projection, while being far more extended in radial direction, their coalescence timescales would be much longer than those for a spherical CG.
In CGs extracted from mock galaxy catalogues built on the outputs of semi-analytical models of galaxy formation run on the Millennium Simulation \citep{springel2005}, a cut on the line-of-sight velocity of 1000 km~s$^{-1}$ from the median line-of-sight velocity of the group members (as also in the sample of \citealp{sohn2015}) increases the fraction of real groups from 35--44 per cent to 59--76 per cent, depending on the applied semi-analytic model (compare CG classes ``mpCG" and ``mvCG" in table 5 of \citealp{millennium}).

The second scenario, long coalescence timescales even for 3D compact groups, is supported by N-body simulations of individual CGs that can survive for up to a few Gyr (e.g. \citealp{barnes1989, governato1991}).
In numerical experiments, the group coalescence time depends on whether the members galaxies share a common halo \citep{mamon1987} and how massive and concentrated their haloes are \citep{athanassoula1997}. 
\citet{tamayo2017} used cosmological simulations to extract the initial positions and velocities of CGs for their \textit{N}-body simulations and argue that CG galaxies fully merge within roughly 5~Gyr.
However, given their lack of information on galaxy magnitudes and (projected) environment, N-body simulations cannot properly evaluate the time span during which the groups would be identified as CG.

Finally, the coalescence timescales of CGs have also been proposed as a test of the underlying cosmological model \citep{kroupa2015, kroupa2016}. In the current standard cosmology ($\Lambda$CDM), each galaxy is embedded in a massive dark matter halo, which is not the case for alternative cosmological models (e.g. MOND, \citealp{milgrom}). The lack of dark matter reduces the dynamical friction and coalescence timescales would therefore be much longer.

Observations are limited to the distribution of galaxies at one given time. Therefore, it is very difficult to obtain an indication of the coalescence timescales of CGs based on observational data. With cosmological simulations, on the other hand, the evolution of galaxy groups can be followed in time and in addition they can provide \mbox{3-dimensional} information on the galaxy configurations. In contrast to the numerical experiments mentioned above, cosmological CGs are embedded in a realistic environment and their haloes evolve self-consistently (i.e. no ideal halo profile as initial condition). 

Cosmological simulations are thus an excellent tool for the study of the nature and time evolution of compact galaxy configurations. Dark matter only cosmological simulations in combination with semi-analytic models of galaxy formation have already been used to show the influence of interlopers of CGs (e.g. \citealp{mcconnachie2008}; \citealp{millennium}) and to indicate that there is no strong link of CGs turning into fossil groups \citep{farhang2017}.
Simulations that also include baryon physics allow a more direct comparison with observations and provide a more realistic combination of orbits, masses, angular momentum (e.g. prograde or retrograde encounter) and luminosities of the member galaxies. 
Using cosmological hydrodynamical simulations allows us to quantify the occurrence of projection effects as well as typical coalescence timescales of CGs in a $\Lambda$CDM universe for a sample of CGs selected with the same criteria as the observed CG sample from \citet{sohn2015}.

We explain the sample selection in Sec.~\ref{methods} and the properties of EAGLE CGs in Sec.~\ref{sec:results}. 
The conclusions and the limitations of the simulations that are relevant for this work are discussed in Sec.~\ref{discussion}. 

In this work, unless explicitly stated otherwise, the coalescence time(scale) or coalescence rate refers to the complete merger of all CG members into one massive galaxy.

\section{Methods} % Main chapter title
\label{methods} % For referencing the chapter elsewhere, use~\ref{Chapter1}

We identify CGs within a cosmological box from the EAGLE (Evolution and Assembly of GaLaxies and their Environments) project (\citealp{eagle}; \citealp{crain2015}). 
EAGLE is a suite of 
hydrodynamical simulations of a standard $\Lambda$CDM \citep{planck2013} universe ($h = 0.6777$; $\Omega_\mathrm{m} = 0.307$; $\sigma_8 = 0.8288$; see table~1 in \citealp{eagle} for a full list of cosmological parameters). 
As the particles are evolved, their properties are saved for 29 snapshots (moments in time) between redshift 20 and 0.
Between $z=1$ and $z=0$ the time between snapshots is 0.61--1.35~Gyr. 

EAGLE is calibrated to match the galaxy stellar mass function and galaxy sizes at $z=0.1$, but has been shown to reproduce a large number of observables over a large range of redshifts, such as the evolution of stellar mass \citep{furlong2015}, galaxy sizes \citep{furlong2017} and the evolution of galaxy angular momentum \citep{swinbank2017}.

The largest EAGLE reference simulation L100N1504, used in this work, has a box size of 100~cMpc (comoving Mpc) and is performed with an equal number of baryonic and dark matter particles ($2\times1504^3$).
The simulation has a force resolution of 0.7~pkpc (proper kpc) which was chosen such that the Jeans scales in the warm (${T\sim10^4}$~K) interstellar medium can be marginally resolved with an initial baryon particle mass of ${m_{\mathrm{g}}=1.8\times10^6\,\mathrm{M}_{\odot}}$ and a dark matter particle mass of ${m_{\mathrm{dm}} = 9.7\times 10^6\,\mathrm{M}_{\odot}}$.

All galaxy properties of the simulated galaxies are extracted from the public EAGLE database \citep{mcalpine2015} using the particles within a 30~pkpc aperture around the centre of the galaxy, defined as the position of the most bound particle.
The SDSS rest-frame magnitudes from \citet{trayford2015} are available for all well-resolved galaxies with stellar mass larger than $M_{\star,\mathrm{min}} = $10$^{8.5}$ M$_\odot$ .

The centre of the cosmological box of each simulation snapshot is placed at a distance corresponding to its redshift (for the snapshot at $z=0$, a redshift distance of $z=0.025$ is used).
We utilise the periodicity of the simulated box and add two additional simulation boxes, one in front and one behind the central simulation box.
The added boxes are especially important for quantifying the fraction of interlopers. 
The observation is simulated using parallel sight lines and the line-of-sight velocities of all galaxies are updated to include the cosmic expansion. 
The apparent \textit{r}-band magnitudes, $m_r$, are calculated from the SDSS rest-frame absolute magnitudes (we use the rest-frame absolute AB magnitudes without dust attenuation) in the EAGLE database for each galaxy depending on its distance to the observer, as well as its Hubble flow velocity (K~correction).
Mock observations are performed from all six sides of the box.

\begin{figure}
\centering
\includegraphics[trim={0 3 30 30},clip,width=0.47\textwidth]{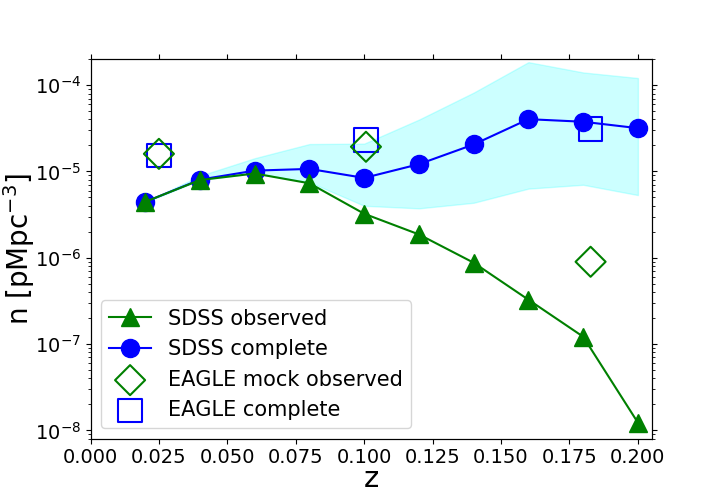}
 \caption{The abundance of CGs as function of redshift in the EAGLE simulation and from the SDSS data as computed by \citet{sohn2015}. The complete sample of the SDSS data corresponds to the completeness correction in \citet{sohn2015} and the complete sample of EAGLE to the sample defined in Sec.~\ref{timeevolution}.
 The errors on the EAGLE abundances based on the spread for mock observations from the 6 different sides are smaller than 0.1~dex and are not shown for visibility.
}
     \label{redshiftabundance}
\end{figure}

For every snapshot, we iterate over all galaxies in the central simulation box (whose centre is located at the snapshot redshift) with $M_{\star} > M_{\star,\mathrm{min}} = $10$^{8.5}$ M$_\odot$ and $14.5<m_r<18$ (as in \citealp{sohn2015}) and search for galaxy configurations that fulfil a slightly modified version of Hickson's criteria \citep{Hickson}, as also used in \citet{sohn2015}:
\begin{enumerate}
\item Population. A CG should have at least 3 galaxy members within 3 magnitudes ($m_{r}$) of the brightest member of the group.
\item Isolation. The angular diameter of the smallest circle enclosing the geometric centres of the group members should be at least 3 times smaller than the angular diameter of the largest circle enclosing no additional galaxies within the magnitude limit of 3 magnitudes or brighter.
\item Compactness. The surface brightness of the compact group should be brighter than 26.0 magnitudes per arcsec$^2$ and is defined as the total magnitude of the galaxy members averaged over the area of the smallest circle enclosing the geometric centres of all members.
\end{enumerate}

Following the method of \citet{sohn2015}, we compute the median line-of-sight velocity of the potential group members and remove the galaxies that differ more than 1500~km s$^{-1}$ from this median value. 
Thereafter, we compute the mean line-of-sight velocity of the remaining galaxies and discard the galaxies that deviate more than 1000~km~s$^{-1}$ from this mean velocity. 
\citet{sohn2015} based this value on the radial velocity restriction implemented in earlier CG studies (since \citealp{hickson1992}). 
The galaxy configuration is identified as a CG if at least three members meet the restrictions on the line-of-sight velocity. As in 
\citet{mcconnachie}, the projected group radius is limited to 1$^{\circ}$.

\begin{table}
\centering
\caption{Median values of selected properties of the EAGLE (`EAGLE mock observed' in Fig.~\ref{redshiftabundance}) and SDSS (`SDSS observed' in Fig.~\ref{redshiftabundance}) CG samples for the whole redshift range.  The errors present the first and third quantile.}
\label{median2d}
\renewcommand{\arraystretch}{1.5} %extra space between rows for visuability of the errors
\begin{tabular}{@{}lcr@{}}
\toprule
\multicolumn{1}{c}{}                    & EAGLE & SDSS \\[-0.2cm] &&\citep{sohn2015}\\ \midrule
Projected group size {[}pkpc{]} & 43$_{-15}^{+22}$      & 85$_{-23}^{+30}$   \\
Median projected separation {[}pkpc{]} & 64$_{-19}^{+28}$         & 101$_{-27}^{+34}$   \\
Log(number density {[}pMpc$^{-3}${]})                        & 3.9$_{-0.5}^{+0.6}$       & $3.1_{-0.4}^{+0.4}$  \\
Velocity dispersion {[}km s$^{-1}${]}   & 161$_{-71}^{+107}$     & 205$_{-85}^{+105}$  \\\bottomrule
\end{tabular}
\end{table}
 
\section{Results}\label{sec:results}
We compare the median properties of CGs in EAGLE to the properties of the CGs in SDSS from \citet{sohn2015} in Table~\ref{median2d}.
The internal number density and the velocity dispersion are estimated from projected properties and the line-of-sight velocity using the same definitions as \citet{sohn2015}.
The median projected group size, separation, density, and velocity dispersion of the EAGLE CGs are consistent with observations, although the simulated CGs tend to be more compact compared to the observed ones. This could be induced by the higher fraction of EAGLE groups with only 3 members (94 per cent) compared to the SDSS sample (42 per cent). The difference in the number of CG members between this work and \citet{sohn2015} is further discussed in Sec.~\ref{discussion}.

In Fig.~\ref{redshiftabundance} we compare the abundance of the two different samples between $z=0.025$ and $z=0.2$.
The green data points indicate the abundance from observations in SDSS (triangles) and the mock observed abundance in EAGLE (diamonds).
Because a magnitude-limited sample of galaxies is used to identify CGs (for both the SDSS observation and EAGLE mock observation), the dependence of the apparent magnitude on the redshift induces a redshift bias.
The blue data points in Fig.~\ref{redshiftabundance} indicate the abundances that are corrected for this effect. The circles represent the completeness corrected abundance from \citet{sohn2015}, while the squares indicate the abundances resulting from our own correction.
The latter is obtained by performing a mock observation of the simulation boxes as if they would be observed at redshift $z=0.025$. 

The completeness correction in \citet{sohn2015} follows the method of \citet{barton1996} and assumes an \textit{r}-band luminosity function (LF) with a faint-end slope of $\alpha = -0.98$ (from \citealp{choi2007}). The EAGLE \textit{r}-band LF \citep{trayford2015} is consistent with the slope of $\alpha = -1.26$ measured in the GAMA (Galaxy and Mass Assembly) spectroscopic survey \citep{loveday2012}. The difference between the two LFs is small ($<30$ per cent) at typical CG luminosities ($M_r<-18$).

Summarizing, the sample of CGs we extracted from the EAGLE simulations are slightly more compact than the sample extracted by \citet{sohn2015} from SDSS, they are a factor of 2 to 4 times more abundant, and consist mainly of 3 members instead of 4. Future comparison with cosmological simulations with different box sizes will show if this behaviour is specific to EAGLE, related to the sample selection or the size of the box, or an intrinsic problem of $\Lambda$CDM simulations. 

Despite these differences, the EAGLE CGs reproduce the shape of the evolution of the abundance with redshift of the completeness corrected observed CGs (see Fig.~\ref{redshiftabundance}). This allows us to use this sample to explain how a high abundance of CGs can be maintained with time, even though their compactness (even more compact than observed CGs) could result in short coalescence timescales due to strong dynamical friction. 

\begin{figure}
\centering
  \includegraphics[trim={0 0 0 0},clip,width=0.47\textwidth]{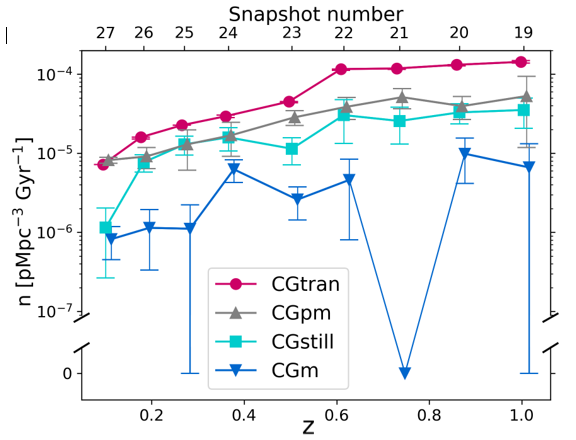}
  \caption{Distribution of the CGs from the completeness corrected sample of EAGLE (defined in Sec.~\ref{timeevolution}) among the different categories based on their identification in the next snapshot. The number densities are normalized for the different time intervals between the snapshots. The points are slightly shifted on the horizontal axis for visibility, but are located at the redshift that corresponds to snapshots 19--27. The errors indicate the standard deviation based on the results from the 6 different observing directions.}
   \label{projvslos}
\end{figure}

 \begin{figure*}
\centering
              \includegraphics[trim={14 29 14 6},clip,width=\linewidth]{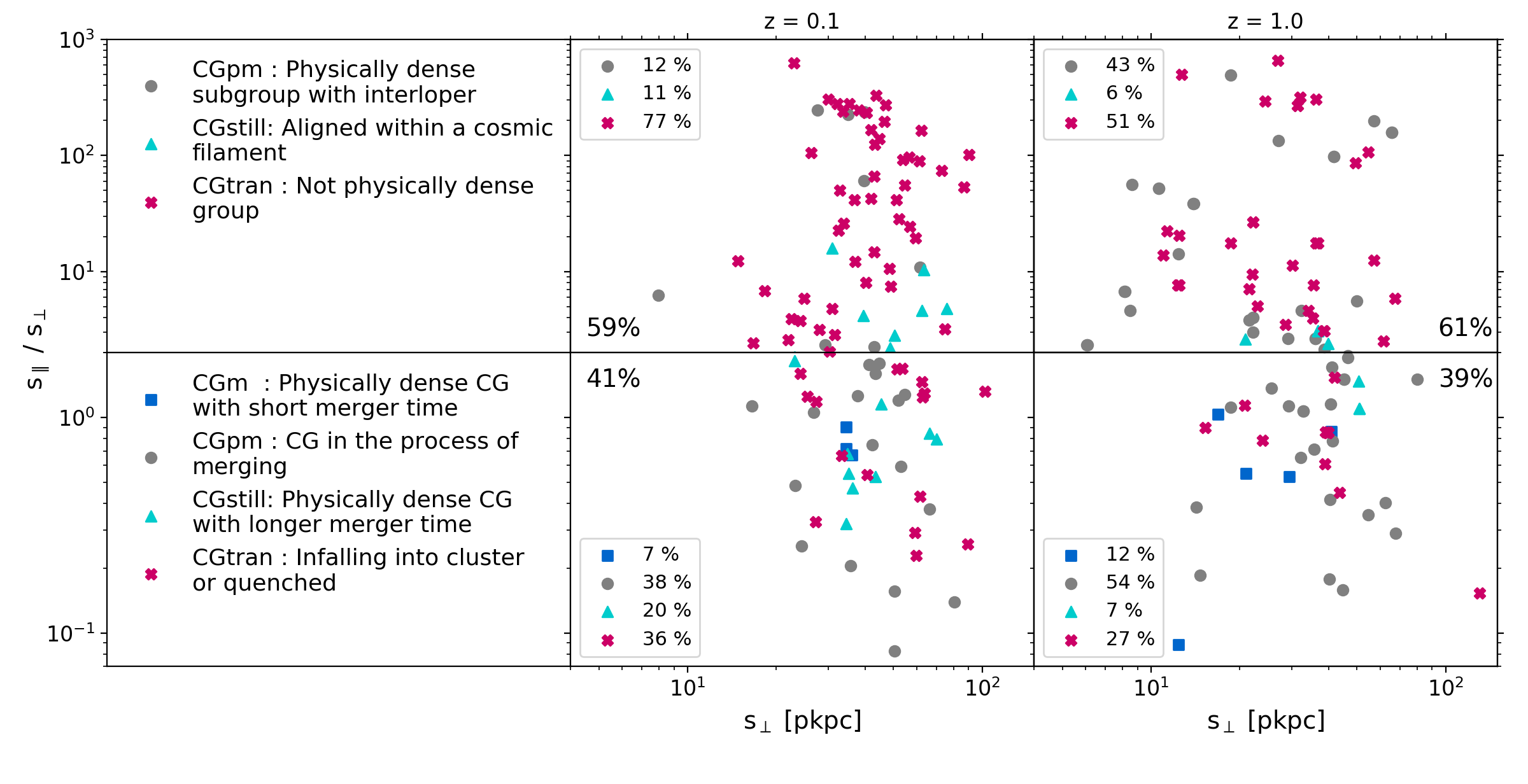}
              \caption{Relative elongation of the CGs along the line of sight ($s_{\parallel} / s_{\perp}$) as a function of the projected group size ($s_{\perp}$) for all EAGLE CGs identified at $z=0.1$ (middle panels) and at $z=1$ (right panels). The bottom panels show groups with $s_{\parallel} / s_{\perp} \leq 2$ while the CGs in the top panels are more elongated along the line of sight ($s_{\parallel} / s_{\perp} \geq 2$). The unboxed percentages represent the distribution between the elongated and physically compact groups. The most likely explanation for each subgroup of CGs (see text for details) in each panel is listed in the left panels. The percentages in the legends indicate the occurrence of the different scenarios per panel.}
               \label{fig:keyfig}
\end{figure*}

\subsection{Time evolution of compact groups}
\label{timeevolution}
Dark matter only simulations of individual CGs have resulted in a wide range of coalescence timescales (from $\ll1$ to a few Gyrs). For a typical coalescence timescale in a cosmological environment, we trace the members of all CGs in the completeness corrected sample (identified between $z=1$ and $0$) forward in time using the merger trees of \mbox{\citet{qu2017}}.
For each galaxy, the EAGLE database contains the ID of its descendant in the next snapshot (for details see \citealp{qu2017}), and a CG is considered fully merged between snapshot $i$ and snapshot $i+1$ if all members of the CG (identified at snapshot $\le i$) share the same descendant.
The group members are identified in the next snapshot and can be used to see how simulated CGs evolve on a time-scale of $\approx$1~Gyr, as for snapshots 19--27 ($z=1$--0.1) the time difference between two snapshots ranges from 0.61 to 1.36~Gyr. After tracing the galaxies one snapshot forward, we distinguish between 4 different scenarios:

\begin{enumerate}
\item \textsl{CGm} (`merged'): all compact group members have merged into one galaxy in the next snapshot.
\item \textsl{CGpm} (`partly merged'): at least two members of the compact group have merged into one galaxy, but not all of them.
\item CGstill (`still a CG in next snapshot'): the compact group members still exist as separate galaxies and are again identified as a compact group in the next snapshot.
\item \textsl{\textsl{CGtran}} (`transient'): the compact group members still exist as separate galaxies, but are not identified as a compact group in the next snapshot anymore.
\end{enumerate}

The occurence of the scenarios per snapshot, normalized for the different time intervals between the snapshots, is visualized in Fig.~\ref{projvslos}. For all redshifts with $z\le 1$, the fractions of CGs that are partly merged (\textsl{CGpm}) or still identified as CGs in the subsequent snapshot (CGstill) are comparable, but only a small fraction (a few per cent) of all CGs fully merge between two snapshots. Most identified CGs are transient (\textsl{\textsl{CGtran}}, i.e. they would only be identified as CGs for a short time). This could be explained if the 2D compactness does not represent the 3D size of the CG (i.e. a projection effect). 

To quantify this effect, we visualize the `roundness' of a group by showing the ratio between the projected group size ($s_\perp$) and the maximum separation between two members along the line of sight (s$_\parallel$) in Fig.~\ref{fig:keyfig}.
We show the distribution of groups among the different scenarios along with suggestions regarding the nature of each subgroup for the lowest ($z=0.1$, left) and highest ($z=1$, right) redshift considered.
We split the identified galaxies into an upper and a bottom panel at $s_\parallel$/$s_\perp=2$, following the criterion for change alignments of \citet{millennium}. 
Of the total sample (snapshots 19--27, $z=1$ to $z=0.1$), 45 per cent of the CGs have small elongations ($s_\parallel$/$s_\perp\le2$) and are therefore physically dense.
The second criterion of \citet{millennium}, a maximum 3D separation smaller than 200h$^{-1}$~kpc, does not exclude additional CGs from being physically dense as this criterion is met for all groups with $s_\parallel$/$s_\perp\le2$.

As expected, most transient CGs (\textsl{\textsl{CGtran}}) have large elongations (top panel) and are therefore only compact in projection. If a CG with a large elongation partially merges (\textsl{CGpm}), it contains a physically close subgroup (e.g. a galaxy pair) plus one or more interlopers.
An interesting class of CGs is the one that contains groups that have a large elongation, but are still identified as a CG in the next snapshot (CGstill in top panel). Observing galaxies along a cosmic filament would provide an explanation and has been suggested in previous work (e.g. \citealp{hernquist1995}). In our sample, this is a minor contribution, in agreement with previous work \citep{millennium}.

Even for physically dense groups ($s_\parallel$/$s_\perp\le2$) the fraction of CGs that merge into one galaxy within 1.35 (0.61) Gyr is only 7 (12) per cent for CGs at $z=0.1$ ($z=1$). 
The large fraction of CGs classified as \textsl{CGpm} and CGstill in the bottom panels of Fig.~\ref{fig:keyfig} already indicates that the typical coalescence timescales of CGs is longer than the time between two snapshots. 
The transient CGs that are compact in 3D ($s_\parallel$/$s_\perp\le2$) lose their classification as a CG because they are either in-falling into a cluster (isolation criterion fails) or have a significant change in brightness (e.g. due to mass stripping) that pushes one of the galaxies out of the magnitude limits\footnote{We verified this for CGs at $z=0.1$ by visualizing the galaxy members as well as their descendants in the subsequent snapshot.} (see Sec.~\ref{methods}).
The labels in Fig.~\ref{fig:keyfig} summarize the above mentioned explanations for each subgroup. 

\begin{figure}
\centering
      \includegraphics[trim={5 24 35 15},clip,width=0.47\textwidth]{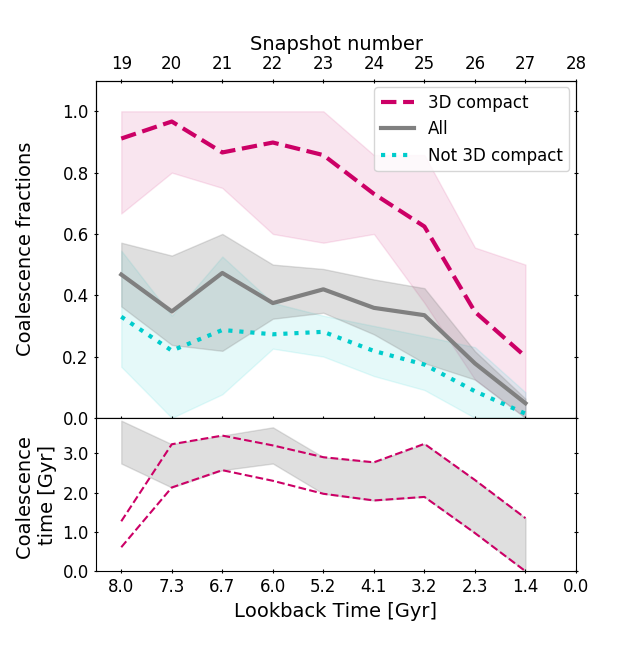}
      \caption{Top: Fraction of CGs whose members merged into one object at or before $z=0$. The full sample (`All', solid line) is split into CGs with $s_{||} / s_{\perp} \le 2$ (dashed line, `3D compact) and $s_{||} / s_{\perp} > 2$ (dotted line; `Not 3D compact'). The shaded region around each line indicates the scatter for the 6 different projections. Bottom: Minimum and maximum (see text) median time between CG identification and full coalescence (`coalescence time') for all (shaded) and 3D compact (dashed lines) CGs.  }
      \label{mergerrate}
\end{figure}

The fractions of CGs that merge into a single object before $z=0$  are visualized in the top panel of Fig.~\ref{mergerrate}.
A roughly constant fraction of $\approx$ 40 per cent of all CGs in our sample (solid line) that are identified with a lookback time from 4 to 8 Gyr merge before $z=0$. 
For the same time span, $\approx 20$ per cent of CGs that are likely to contain a larger fraction of chance projections ($s_{\parallel}/s_{\perp} \geq 2$, dotted line) fully merge, while this fraction increases to $>80$ per cent for CGs that are compact in 3D (dashed line).

Selecting only CGs from our sample that fully merge into one galaxy throughout the simulation (i.e. before $z=0$) allows us to constrain the typical coalescence timescales for CGs in EAGLE. As the full coalescence can occur at any time between the snapshots, the median times for the minimum and maximum coalescence timescales are evaluated separately\footnote{For example: a CG is identified at snapshot 25 and is listed as fully merged at snapshot 28 ($z=0$). The minimum coalescence timescale is the time between snapshot 25 and 27 (1.89 Gyr), while the maximum coalescence timescale is the time between snapshots 25 and 28 (3.24 Gyr).}. The bottom panel of Fig.~\ref{mergerrate} illustrates the range of these median values for all identified CGs within a given snapshot (grey band). We conclude that CGs that merge between $z=1$ and $z=0$ have a typical coalescence time of 2--3~Gyr. This is constant over a large range of the considered lookback times (redshifts) and only decreases at late times, as longer coalescence times are not sampled anymore. 

\section{Discussion}
\label{discussion}

In this work, we constructed a sample of compact groups of galaxies (CGs) from all snapshots with $z\le1$ of the 100 cMpc box of the EAGLE suite of cosmological hydrodynamical simulations. As the stellar distributions within the simulated galaxies are modelled self-consistently, their apparent \textit{r}-band magnitudes can be derived and the same selection criteria as in observed samples of CGs (here: \citealp{sohn2015}) are used. 

This study provides a unique perspective of the nature of CGs in a $\Lambda$CDM universe and shows a few scenarios that together can explain the CG abundance in the EAGLE simulation. 
Compared to observed CGs, the EAGLE CGs are slightly more compact, more abundant and lack CGs with 4 members.
Future studies will reveal if these differences are related to the selection criteria, intrinsic galaxy properties, or cosmic variance (see below).
For this study, we focused on explaining the abundance of CGs for $z<0.2$ in a $\Lambda$CDM universe by a combination of projection effects and long coalescence timescales. The shape of the evolution of the abundance of CGs with time is well reproduced by the EAGLE CGs.

Comparing the radial extent of CGs to their projected size (Fig.~\ref{fig:keyfig}) illustrates the importance of projection effects, even if a line-of-sight velocity of 1000~km~s$^{-1}$ is applied for the sample selection. 
59 (61) per cent of CGs at $z=0.1$ ($z=1$) have large elongations and are therefore not compact in 3D, only in projection.
These change alignments typically last for ${< 1}$~Gyr, and are continually replaced by new projected CGs. 

Numerical estimates of the coalescence timescales of CGs based on simulations of individual dark matter haloes depend on the initial conditions for position, velocities and halo shapes. Therefore, simulated CGs can merge either very fast (${\ll 1}$~Gyr) or on longer (${\gg 1}$~Gyr) timescales. The CGs in EAGLE emerge naturally from the large-scale structure and therefore show a realistic distribution of orbits, masses, and halo shapes. This allows us to constrain the typical (i.e. median) coalescence timescale in our sample of CGs. While a small percentage of CGs has merged into a single object at the next snapshot (1.35 Gyr at $z=0.1$, 0.61 Gyr at $z=1$), the typical coalescence timescale is longer (2--3~Gyr).

In combination with projection effects, we showed that a typical coalescence timescale of $>1$~Gyr in $\Lambda$CDM can explain the evolution of the CG abundance in redshift as identified in observations for $z<0.2$ 

Using hydrodynamical simulations for the study of CGs brings a lot of advantages, but limitations in volume (of the simulated box) as well as resolution affect our CG sample.
Hydrodynamical simulations are computationally more expensive than dark matter only simulations. The state-of-the-art EAGLE simulation box with a volume of (100 cMpc)$^3$ is therefore 400 times smaller than the dark matter only Millennium simulation \citep{springel2005} which is used by e.g. \citet{mcconnachie2008} and \citet{millennium} to construct simulated CG catalogues.
Rare objects, such as massive haloes and massive galaxies, are not fully sampled in small volumes and cosmic variance can affect the clustering signal also on the smallest scales, if the signal is dominated by satellite galaxies of individual high mass haloes (as shown for red EAGLE galaxies in \citealp{artale2017}). 

Furthermore, the faint-end slope of the luminosity function within a halo (conditional luminosity function) seems to steepen with increasing halo mass in SDSS galaxies (\citealp{yang2008}, see also \citealp{trevisan2017}). If more massive haloes had a higher probability of containing richer CGs (i.e. CGs with more members) or additional interlopers, simulations of limited volumes can be biased. As reference, the median parent halo mass of EAGLE CGs at $z=0.1$ that share the same halo is $\langle M_{\mathrm{200,crit}} \rangle =1.3 \times 10^{12}\,\mathrm{M}_{\odot}$\footnote{The halo mass $M_{\mathrm{200,crit}}$ is defined as the mass that encloses a mean density of 200 times the critical density of the Universe.} and only 15 per cent of EAGLE CGs are in haloes with $M_{\mathrm{200,crit}}>10^{13}\,\mathrm{M}_{\odot}$. 

In addition to the limitations in volume, if the projected distance between individual members is very small (as typical for CGs) both observations and simulations are prone to missing group members in different ways. While redshift dependent fiber collision and blending (see discussions in \citealp{millennium} and \citealp{2016sohn}) can occur in observations, the exact identification of close pair members (major mergers) depends on the substructure finder used in simulations (e.g. \citealp{behroozi2015}).
Creating complete mock observations of galaxies, including a full light cone, and sending these images through the same pipeline that is used to construct the SDSS samples of compact groups of galaxies will overcome these issues in future work.

The combination of cosmic variance and systematic differences between the different limitations in observations and simulations might contribute to the lack of 4-member CGs in the EAGLE CG sample, compared to the sample from \citet{sohn2015}. It is worth mentioning that with a slightly updated CG identification algorithm and a larger galaxy sample (SDSS data release 12), \citet{2016sohn} also found a significantly lower fraction of 4-member groups than \citet{sohn2015} with SDSS data release 6. They found more than 5 times as many 3-member CGs than CGs with $\ge4$ members, which is much closer to the few percent of CGs with $\ge4$ members from this work.

While we cannot exclude that additional members with stellar masses below $M_{\mathrm{min}}$ could be missed, the majority of CG members in EAGLE are well above the resolution limit. For example, at $z=0.1$ the median stellar mass of CG members is $\log M_{\star} [\mathrm{M}_{\odot}] = 9.4$ ($\approx 1400$ star particles), which is 0.4 dex higher than the median stellar mass of non-CG members (with magnitude entries) and 0.9 dex above the minimum stellar stellar mass $\log M_{\star} [\mathrm{M}_{\odot}] = 8.5$. 

Despite these limitations, as the used simulation box (EAGLE RefL0100N1504) reproduces the \textit{r}-band luminosity function \citep{trayford2015}, the observed clustering of galaxies \citep{artale2017}, as well as the stellar mass - halo mass relation \citep{eagle}, the EAGLE CGs are expected to be a realistic (albeit potentially biased) representation of CGs in the observed Universe.
The individual processes identified in this work that explain the abundance of CGs in EAGLE can therefore still be considered realistic contributions to the observed abundance of CGs.

The sample of EAGLE CGs that we have used is public\footnote{https://www.sylviaploeckinger.com/supplementary-material}, so future research can be conducted easily from the same sample.

\section*{Acknowledgements}
We would like to thank Gary Mamon for his comprehensive and insightful comments during the refereeing process.
SP was supported by European Research Council (ERC) Advanced Investigator grant DMIDAS (GA 786910, PI C. S. Frenk). \\ 

\DeclareRobustCommand{\VAN}[3]{#3}

\bibliographystyle{mnras}
\bibliography{paper_draft}

\appendix

% Don't change these lines
\bsp	% typesetting comment
\label{lastpage}
\end{document}